\documentclass[usenatbib,usegraphicx]{mn2e}

\newcommand{\der}[2]{\ensuremath{\frac{d #1}{d #2}}}

\newcommand{\ie}{{\it i.e.}}

\newcommand{\eg}{{\it e.g.}}

\newcommand{\be}{\begin{equation}}
\newcommand{\ee}{\end{equation}}
\newcommand{\bea}{\begin{eqnarray}}
\newcommand{\eea}{\end{eqnarray}}

\title[Structure of relativistic reconfinement shocks]
{A structure and energy dissipation efficiency of relativistic reconfinement shocks}

\author[K. Nalewajko \& M. Sikora]
{Krzysztof Nalewajko\thanks{E-mail: knalew@camk.edu.pl} and Marek Sikora \\
Nicolaus Copernicus Astronomical Center, Bartycka 18, 00-716 Warsaw, Poland}

\begin{document}

\maketitle

\begin{abstract}

We present a semi-analytical hydrodynamical model for the structure of  
reconfinement shocks formed in astrophysical relativistic jets interacting
with external medium. We take into account exact conservation laws, both
across the shock front and in the zone of the shocked matter, and exact 
angular relations. Our results confirm a good accuracy of the 
approximate 
formulae derived by Komissarov \& Falle (1997). However, including
the transverse pressure gradient in the shocked jet, we predict an absolute
size of the shock to be about about twice larger.
We calculate the efficiency of the kinetic energy dissipation in the shock
and show a strong dependence on both the bulk Lorentz factor
 and opening angle of the jet.
\end{abstract}


\section{Introduction}

Geometry (cross-sectional size, opening angle, and bending) of 
supersonical, light jets
is regulated, in general, by a complex system of oblique shocks. 
At certain circumstances
they take form of reconfinement shocks \citep{1983ApJ...266...73S}. 
Such shocks
have been considered to be responsible for non-thermal activity  
in AGN radio cores
(see, \eg, \citealt{1988ApJ...334..539D}; \citealt{1997MNRAS.288..833K} -- 
hereafter KF97;
\citealt{2006MNRAS.370..981S}) and, on much larger distances, in  
kiloparsec-scale 
radio knots \citep{1998MNRAS.297.1087K}. They are also predicted to 
operate
in massive X-ray binary systems \citep{2008A&A...482..917P} and in
GRB collapsars (\citealt{2007ApJ...671..678B}).

A direct way to verify whether a given source can be interpreted
in terms of a reconfinement shock is to determine whether location
and extension of the source is consistent with a power of a jet and
pressure/density of external medium. Under several simplifying assumptions
analytical formulae relating these quantities were derived by
\cite{1991MNRAS.250..581F} and, for relativistic shocks, by KF97.
We have developed a semi-analytical model based on exact conservation laws
and an exact dependence of a shock structure on an initial opening angle of a jet.
Like in KF97, we adopt the cold jet approximation, \ie~we neglect the internal 
energy of the unshocked jet matter.
The aim of this paper is to test the accuracy of the analytical formulae
and to study the effects of including a transverse pressure gradients
in the post-shock zone.
%One of our models takes into account the pressure gradients in the post-shock zone.

Our models are described in \S2. They are compared with analytical results of
KF97 in \S3. We study the efficiency of energy dissipation in the reconfinement
shocks in \S4 and discuss their possible 'astrophysical appearance'
in \S5. Our main results are summarized in \S6.

\section{Description of the reconfinement models}

The models we develop here are stationary, axisymmetric and purely hydrodynamical.
We use a cylindrical coordinate system originating at the central source,
with z-axis aligned with the jet symmetry axis and $r$ denoting the cylindrical radius.
At every point, the flow is characterised by the following parameters: bulk Lorentz factor
$\Gamma=1/\sqrt{1-\beta^2}$, rest-density $\rho$, pressure $p$ and the angle between
the velocity vector and the z-axis $\theta$. We use the equation of state for the ideal gas
%after \citet{Sy57}:
%
\be
\label{eqst1}
p=(\gamma-1) e\,,
\ee
where $e$ is the internal energy density and $\gamma$ is the numerical coefficient,
which for non-relativistic and for ultra-relativistic gases coincides
with the adiabatic index with a value $5/3$ and $4/3$, respectively.
For intermediate cases and/or mixtures of non-relativistic and relativistic gases
$\gamma$ takes intermediate values which might somewhat differ from the respective
values of the adiabatic indices (see KF97).

In a stationary flow conservation of mass, energy and momentum is expressed by 
the following equations:
\bea
\label{eqcon1}
\partial_i(\rho u^i) &=& 0\,,\\
\label{eqcon2}
\partial_iT^{i\mu} &=& 0\,;
\eea
where
\be
T^{\mu\nu}=wu^\mu u^\nu+pg^{\mu\nu}
\ee
is the energy-momentum tensor,
\be
w= \rho c^2+e+p
\ee
is the enthalpy, $u^\mu$ is the 4-velocity (its spacial components are 
$u^i=\Gamma \beta^i$),  and 
$g^{\mu\nu}$ is the metric tensor of signature ($-+++$). For the purpose of this 
study we assume a flat Minkowski space.

The jet is modelled as a spherically symmetric adiabatic outflow from the central source
into the cone of half-opening angle $\Theta_j$. Equations (\ref{eqcon1} -- \ref{eqcon2})
can be used to show that mass and energy fluxes of the upstream flow
through any given solid angle must be conserved:
\bea
\label{eqcon3}
\der{}{R}\left[R^2(\rho_ju_j)\right] &=& 0\,,\\
\label{eqcon4}
\der{}{R}\left[R^2(\Gamma_jw_ju_j)\right] &=& 0\,;
\eea
where $R=\sqrt{r^2+z^2}$ is the radial distance from the central source. 
The quantity $\Gamma_jw_j/\rho_j$ is invariant along $R$. 

In the cold jet approximation $p_j$ is negligible, so $w_j\simeq\rho_jc^2$.
Then $\Gamma_j$ is invariant along $R$, so from equation (\ref{eqcon4}) we have
$w_j\propto R^{-2}$. The total power of a jet is
%$L_j$ may be used to calculate enthalpy at given $R$:
%
\be
\label{eqlj}
L_j=w_j\Gamma_ju_jc\times 2\pi(1-\cos\Theta_j)R^2\,.
\ee
Given $\Theta_j$, $L_j$ and $\Gamma_j$, it is now possible to 
calculate the flow parameters
for every point within the jet.

\begin{figure}
\includegraphics[width=\columnwidth]{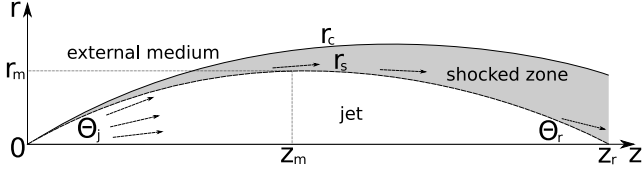}
\caption{Structure of the reconfinement shock for static external medium.}
\label{fig1}
\end{figure}

In the interaction between the a jet and external matter a double oblique shock 
structure forms, but when external medium is static, it degenerates into 
a single shocked zone (see Fig. \ref{fig1}). In this scheme, the jet is bounded
 by the inner shock surface $r_s (z)$ and the shocked gas is bounded by 
the contact discontinuity $r_c(z)$. We denote the inclination angles of these 
surfaces by $\tan\alpha_{s(c)}=dr_{s(c)}/dz$, respectively.
The flow parameters ($\Gamma$, $\rho$, $p$ and $\theta$)
are marked with the following subscripts: $j$ -- for the jet matter at $r_s$, 
$s$ -- for the shocked matter at $r_s$, $c$ -- for the shocked matter at $r_c$ 
and $e$ -- for external medium at $r_c$.

The conservation laws must be satisfied across the shock surface.
Equations (\ref{eqcon1} -- \ref{eqcon2}) can be used to obtain shock jump conditions
\citep{1959flme.book.....L}:
\bea
\left[\beta_\parallel\right] &=& 0\,,\\
\left[\rho u_\perp\right] &=& 0\,,\\
\left[wu_\perp^2+p\right] &=& 0\,,\\
\left[\Gamma wu_\perp\right] &=& 0\,;
\eea
where $u_\parallel$ and $u_\perp$ are the tangent and normal components
(with respect to the shock surface) of the local velocity field, respectively. 
At the shock front $r_s$, they give:
\bea
\label{eq1a}
\beta_s\cos(\theta_s-\alpha_s)&=&\beta_j\cos(\theta_j-\alpha_s)\,,\\
\label{eq1b}
u_s\rho_s\sin(\theta_s-\alpha_s)&=&u_j\rho_j\sin(\theta_j-\alpha_s)\,,\\
\label{eq1c}
u_s^2w_s\sin^2(\theta_s-\alpha_s)+p_s&=&u_j^2w_j\sin^2(\theta_j-\alpha_s)+p_j\,,\\
\label{eq1d}
\Gamma_su_sw_s\sin(\theta_s-\alpha_s)&=&\Gamma_ju_jw_j\sin(\theta_j-\alpha_s)\,.
\eea
For the contact discontinuity there must be no flow through the surface, so the
constraints derived from equations (\ref{eqcon1} -- \ref{eqcon2}) are much more simple:
\bea
\label{eq2a}
p_c &=& p_e\,,\\
\theta_c &=& \alpha_c\,.
\eea
The purpose of our models is to find the geometrical structure and
the physical conditions of the shocked zone,
given the conditions in the jet and in the external medium.
%The $\gamma$ parameter must be assumed a'priori for each zone, but for
%adiabatic flows it only ranges from $4/3$ (ultra-relativistic matter) to $5/3$
% (non-relativistic matter).

\subsection{Model 1}

In our first model we adopt an assumption made by \citep{2007ApJ...671..678B},
that the shocked zone
has no transverse structure, which means that for a given $z$: $\Gamma_c=\Gamma_s$,
$\rho_c=\rho_s$, $p_c=p_s$, $\theta_c=\theta_s$. Knowing that $p_s=p_e$, we can solve
equations (\ref{eq1a} -- \ref{eq1d}) for the 4 unknowns: $\Gamma_s$, $\rho_s$,
$\theta_s$, $\alpha_s$. This can be done explicitly using exact analytical formulae
(see Appendix \ref{appendix1}). We may then find $r_s$ by numerical integration over $z$.

We considered also finding the contact discontinuity surface $r_c$, 
by noting that 
$\alpha_c=\theta_s$. But when we calculated the mass flux across the shocked 
zone, we found that it is not conserved. Moreover, when neglecting the 
transverse pressure gradients, one cannot satisfy the transverse momentum 
balance needed to account for the curvature of the streamlines.

\subsection{Model 2}

A simple transverse structure of the shocked zone can be included in our model by
assuming that parameters at opposite boundaries are independent. For a given $z$ we now
have 4 more unknown parameters: $\Gamma_c$, $\rho_c$, $\alpha_c$ and $p_s$.
Therefore we need 4 additional equations that would tie the flow parameters at the shock surface to the flow parameters at the contact discontinuity.

We use conservation laws across the shocked zone introduced by
\cite{2007ApJ...671..678B}:
\be
\label{eq3a}
\der{}{z}\left[\int_{r_s}^{r_c}u\rho\cos\theta r\;dr\right]+\rho_su_s^kn_{sk}\frac{r_s}{\cos\alpha_s} =0 \,,
\ee
\be
\label{eq3b}
\der{}{z}\left[\int_{r_s}^{r_c}T^{\mu z}r\;dr\right]+T_s^{\mu k}n_{sk}\frac{r_s}{\cos\alpha_s}+T_c^{\mu k}n_{ck}\frac{r_c}{\cos\alpha_c}=0;
\ee
where $\vec{n}_s$ and $\vec{n}_c$ are the vectors normal to the shock surface and
contact discontinuity surface, respectively, oriented outwards the shocked zone.
Equation (\ref{eq3a}) describes conservation of mass, while the equation set (\ref{eq3b})
includes conservation of energy ($\mu=0$) and of two momentum components
(radial -- $\mu=r$, longitudinal -- $\mu=z$). Note that
$T_c^{ik}n_{ck}= p_cn_c^i \ne 0$. We can solve these equations by reducing
them to a system of linear ODEs (see Appendix \ref{appendix2}).

\section{Geometrical properties of reconfinement shocks}

The crucial characteristic of the reconfinement shocks is their length 
scale, which may be estimated observationally. KF97
provided simple analytic formulae, in which they connect geometrical properties
of the shock surface to physical parameters, such 
as the external pressure $p_e$,
the total power of a jet $L_j$, and its bulk Lorentz factor $\Gamma_j$.
They assumed that: the pre-shock plasma is cold ($p_j \ll \rho c^2$);
pressure behind the shock is equal to the external pressure ($p_s=p_e$);
the half-opening angle $\Theta_j$ is small; and the pressure balance at the shock front,
given by equation (\ref{eq1c}), can be approximated by the formula:
\be
p_s=\mu u_j^2\rho_jc^2\sin^2(\theta_j-\alpha_s)\,,
\ee
with $\mu=17/24$.

Below, we rewrite their results, using slightly different notation. We launch the jet
from the distance $z=z_0$. Let the external pressure profile be
$p_e(z)=p_0(z/z_0)^{-\eta}$ (we expect $\eta\ge 0$). The shock surface should satisfy
a boundary condition $r_s(z_0)=z_0\tan\Theta_j$, where $\Theta_j$ is the jet
half-opening angle. Then the shock surface equation is:
\be
r_s(z)=\left[1-\frac{z_0^{\eta/2}}{\delta\Lambda}\left(z^\delta-z_0^\delta\right)\right]\Theta_jz\,,
\ee
where $\delta=1-\eta/2$, and 
\be
\label{eq3.1}
\Lambda=\sqrt\frac{\mu\beta_jL_j}{\pi p_0c}\,,
\ee
is a characteristic length scale\footnote{
Note that in the Eq. (\ref{eq3.1}) we have $\Lambda \propto L_j^{1/2}\beta_j^{1/2}$, while
models of nonrelativistic jets predict $\Lambda \propto \tilde{L}_j^{1/2}\beta_j^{-1/2}$
(see Eq. (1) in \cite{1994MNRAS.266..649K} and refs. therein).
The reason for the difference is that the term $L_j$ includes
the flux of the rest energy, $\dot M_j c^2$, while $\tilde{L}_j$ doesn't.}.
The reconfinement is found for $\eta<2(1+z_0/\Lambda)$ at
\be
z_r=z_0\left(1+\delta\frac{\Lambda}{z_0}\right)^{1/\delta}\,.
\ee
The maximum width of unshocked jet,
\be
r_m=\frac{z_0^2}{\Lambda}\left(\frac{z_r}{(1+\delta)z_0}\right)^{1+\delta}\Theta_{j}\,,
\ee
is achieved at
\be
z_m=\frac{z_r}{(1+\delta)^{1/\delta}}\,.
\ee
The aspect ratio of the jet is given by
\be
\frac{r_m}{z_r}=\frac{\delta+\frac{z_0}{\Lambda}}{(1+\delta)^{1+\delta}}\Theta_{j}\,.
\ee
The shock surface inclination at $z_0$ is
\be
\Theta_0=\Theta_j\left(1-\frac{z_0}{\Lambda}\right)\,.
\ee
The half-closing angle (equal to minus the shock inclination at $z_r$) is
\be
\Theta_r=\Theta_j\left(\delta+\frac{z_0}{\Lambda}\right)\,.
\ee

\begin{figure}
\includegraphics[width=\columnwidth]{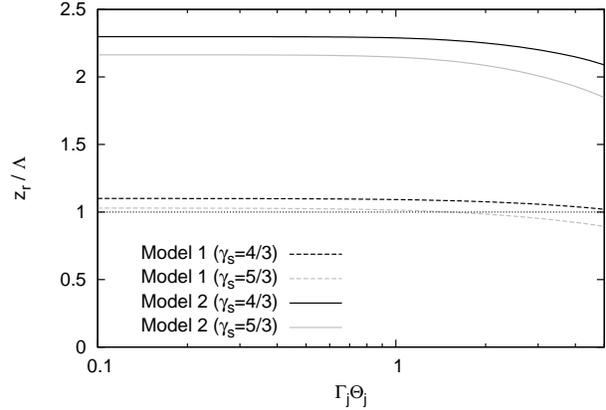}
\caption{The ratio of reconfinement position $z_r$ to the characteristic length
$\Lambda$ as a function of $\Gamma_j\Theta_j$. Results for Model 1
(\emph{dashed lines}) and Model 2 (\emph{solid lines}) are shown for
different equations of state of the shocked matter: $\gamma_s=4/3$
(\emph{black lines}) and $\gamma_s=5/3$ (\emph{grey lines}).}
\label{fig2}
\end{figure}

\begin{figure}
\includegraphics[width=\columnwidth]{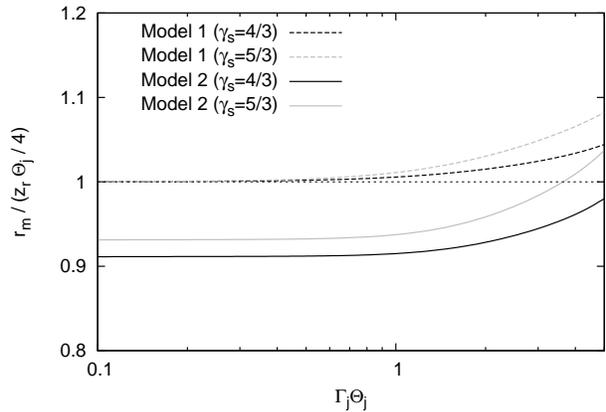}
\caption{The aspect ratio of unshocked jet $r_m / z_r$, divided by the value
$\Theta_j / 4$ predicted by KF97, as a function of $\Gamma_j\Theta_j$.
The linestyles are the same as in Fig. \ref{fig2}.}
\label{fig3}
\end{figure}

\begin{figure}
\includegraphics[width=\columnwidth]{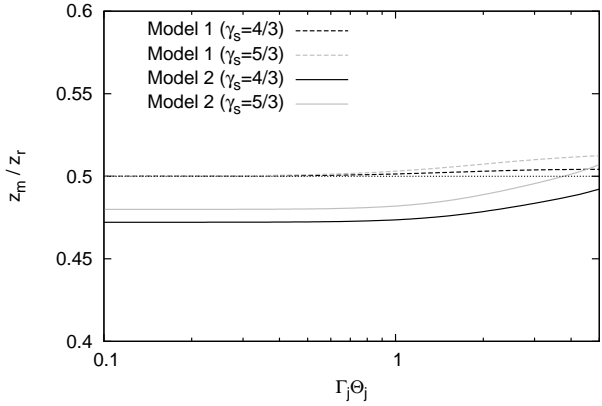}
\caption{The ratio of the maximum jet width position $z_m$ to the reconfinement
position $z_r$ as a function of $\Gamma_j\Theta_j$. The linestyles are the same
as in Fig. \ref{fig2}.}
\label{fig4}
\end{figure}

\begin{figure}
\includegraphics[width=\columnwidth]{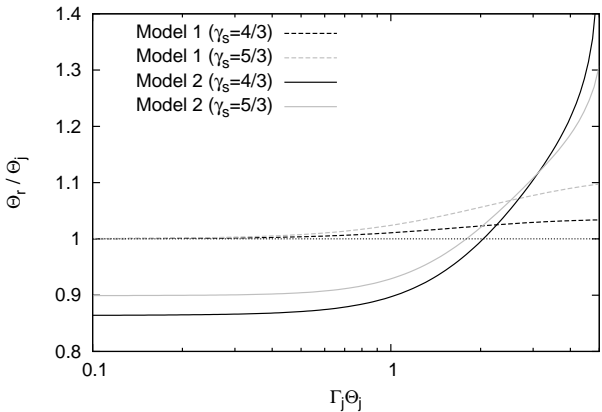}
\caption{The ratio of the half-closing angle $\Theta_r$ to the half-opening
angle $\Theta_j$ as a function of $\Gamma_j\Theta_j$. The linestyles are the same
as in Fig. \ref{fig2}.}
\label{fig5}
\end{figure}

For the case of uniform external pressure ($\eta=0$, $\delta=1$) and a jet
originating close to the central source ($z_0\ll\Lambda$) we find the shock
to be parabolic, with very simple characteristics: $z_r\simeq\Lambda$,
$r_m/z_r\simeq\Theta_j/4$, $z_m\simeq z_r/2$,
$\Theta_0\simeq\Theta_r\simeq\Theta_j$. We have tested these
relations in our two models, the results are shown on Figs. \ref{fig2} -- \ref{fig5},
as a function of half-opening angle $\Theta_j$ for a fixed $\Gamma_j=10$.
Other parameters used were $L_j=10^{46}\;\rm erg\cdot s^{-1}$,
$p_0=10^{-2}\;\rm dyn$ and $z_0=10^{15}\;\rm cm$. The characteristic length
for these parameters is $\Lambda=2.74\cdot 10^{18}\;{\rm cm}=0.89\;{\rm pc}$.

A very good agreement between the results of Model 1 and
the analytical formulae results from the same
value of the pressure behind the shock ($p_s=p_e$). Deviations for
$\Gamma_j\Theta_j>1$ reflect the small angle approximation employed in
analytical formulae. Small but systematic deviations of $z_r$ from $\Lambda$
result from approximate pressure balance equation.

In Model 2 the pressure behind the shock is systematically lower than $p_e$,
but it cannot be fitted to a single power-law function of $z$. This results
in longer reconfinement structures (by a factor of about $2.2$). We have found
that for small and intermediate half-opening angles: $r_m/z_r < \Theta_j/4$, $z_m<z_r/2$ and, accordingly, $\Theta_r<\Theta_j$. For large angles
the deviations from analytical predictions are more pronounced.
Nevertheless, the effects of independent values for the $p_s$ are not
particularly strong. The analytical formulae are still very useful within
the order of magnitude accuracy.

\section{Energy dissipation}

\begin{figure}
\includegraphics[width=\columnwidth]{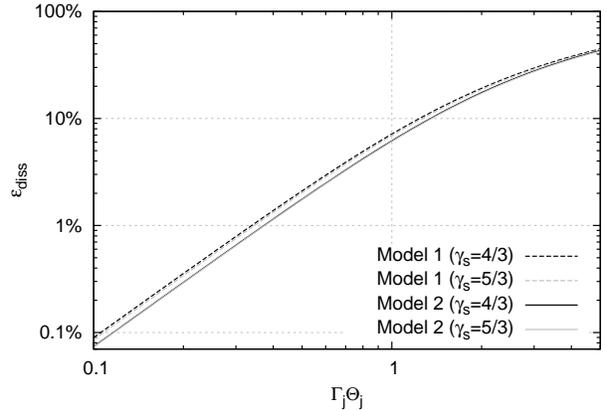}
\caption{Dissipation efficiency $\epsilon_{diss}$ as a function of $\Gamma_j\Theta_j$,
calculated for $\Gamma_j=10$. The linestyles are the same as in Fig. \ref{fig2}.}
\label{fig6}
\end{figure}

The kinetic energy flux through the shock front is dissipated with efficiency
\be
\label{eqdiss1}
\epsilon_{diss}\equiv \frac{F_{kin(j)} - F_{kin(s)}}{F_{kin(j)}}
\ee
where 
\be
\label{eqdiss2}
F_{kin} \equiv \rho c^2 u_\perp (\Gamma - 1)
\ee 
and $u_\perp$ is the 4-velocity component normal to the shock front.
Combining Eqs. (\ref{eqdiss1}), (\ref{eqdiss2}) and (\ref{eq1b}) gives
\be \epsilon_{diss} =
%\frac{\rho_ju_{j \perp}(\Gamma_j-1) - \rho_su_{s \perp}(\Gamma_s-1)}{\rho_ju_{j \perp}(\Gamma_j-1)} =
\frac{\Gamma_j - \Gamma_s}{\Gamma_j-1}\,.
\ee
As averaged over the entire shock front area, the efficiency of energy 
dissipation is found to strongly depend on the product  $\Gamma_j\Theta_j$.
Results for both models with a fixed $\Gamma_j=10$ are shown 
in Fig. \ref{fig6}. We find that the averaged efficiency is very similar 
in both models and is insensitive to the value of $\gamma_s$. 
It approximately scales like $\epsilon_{diss}\sim 0.06 (\Gamma_j\Theta_j)^2$
for $\Gamma_j\Theta_j <1$,
but its increase slows down at $\Gamma_j\Theta_j >1$.
% but saturates for $\Gamma_j\Theta_j>1$.}

\begin{figure}
\includegraphics[width=\columnwidth]{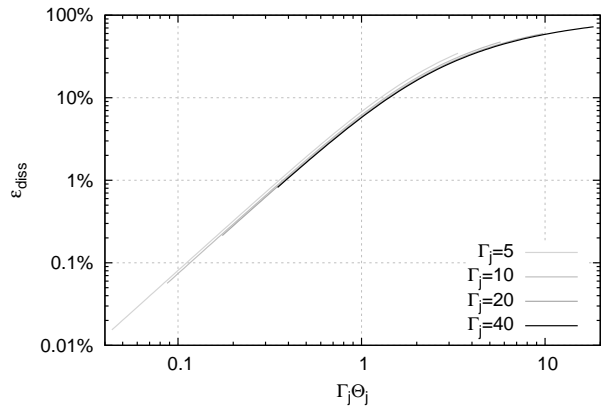}
\caption{Dissipation efficiency $\epsilon_{diss}$ as a function 
of $\Gamma_j\Theta_j$, calculated for Model 2 with $\gamma_s=4/3$.  
Line colour indicates the value of $\Gamma_j$: 5 (\emph{light grey}), 
10 (\emph{grey}), 20 (\emph{dark grey}) and 40 (\emph{black}).}
\label{fig7}
\end{figure}

In order to determine whether $\epsilon_{diss}$ is truly
a function of $\Gamma_j\Theta_j$, in Fig. \ref{fig7} we present the results 
for Model 2 with $\gamma_s=4/3$ and different values of $\Gamma_j$. We find little 
discrepancy between the curves, which implies that it is a well defined dependence.

\begin{figure}
\includegraphics[width=\columnwidth]{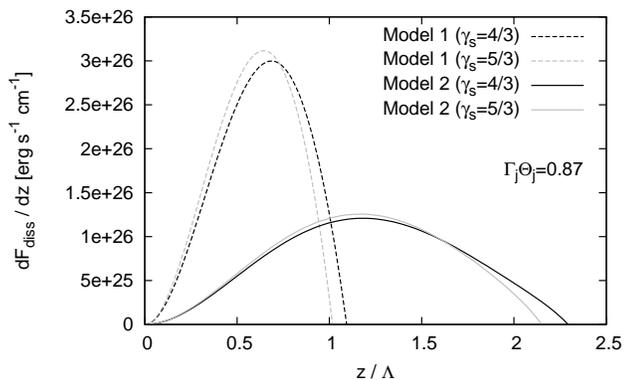}
\caption{Profiles of dissipated energy flux produced at the shock surface, calculated for $\Gamma_j=10$ and $\Theta_j=5^\circ$. The linestyles are the same as in Fig. \ref{fig2}.}
\label{fig8}
\end{figure}

\begin{figure}
\includegraphics[width=\columnwidth]{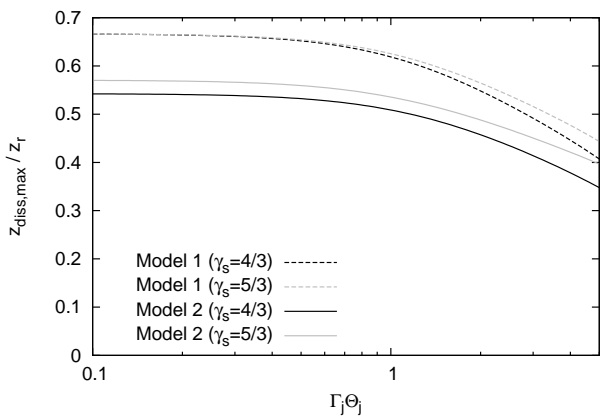}
\caption{Ratio of the location of maximum of dissipated energy $z_{diss,max}$ to the reconfinement position $z_r$ as a function of $\Gamma_j\Theta_j$. The linestyles are the same as in Fig. \ref{fig2}.}
\label{fig9}
\end{figure}

We have investigated the $z$-profiles of the dissipated energy flux. 
In Fig. \ref{fig8} we show the results for both models, with $\Gamma_j=10$ and $\Theta_j=5^\circ$. Although the reconfinement position $z_r$ is 
more than twice large in Model 2, as compared to Model 1, 
the total amount of dissipated energy is very similar. The dissipated
energy profiles have a well defined maximum, which we denote 
as $z_{diss,max}$.  The ratio of $z_{diss,max}$ to $z_r$ is shown 
in Fig. \ref{fig9}. It is larger in Model 1, but larger than $1/2$ 
in both models for $\Gamma_j\Theta_j<1$. It decreases strongly with 
increasing $\Gamma_j\Theta_j$, for $\Gamma_j\Theta_j>1$.

Noticing that the energy flux
$F_w = w \Gamma u_\perp = (\rho c^2+\gamma e)\Gamma u_\perp$ is conserved
across the shock front (see Eq. \ref{eq1d}), one can find that,
for $\gamma_j\sim\gamma_s$, the efficiency of the internal energy production is
\be
\epsilon_e \equiv \frac{F_{e(s)}-F_{e(j)}}{F_{kin(j)}} \sim \frac{1}{\gamma_s}\epsilon_{diss}\,.
\ee
A fraction of this energy
is tapped by particles accelerated to relativistic energies and lost by 
nonthermal radiation. Such processes, if efficient, may significantly affect
the shock structure.

%In an extreme case, if all dissipated energy is
%immediately lost, no pressure develops between the shock front '$r_s$'
%and discontinuity '$r_c$' and surfaces of  both converge into one surface,
%with all the immediately cooled post-shock matter moving on this surface. 
%Geometry of this surface is equivalent to the surface '$r_s$' given by 
%our Model 1. Since the structure of the shock calculated by our Model 2
%is calculated assuming no energy losses (adiabatic case), '$r_s$' solutions 
%for models with  intermediate radiative losses are expected 
%to be enclosed between '$r_s$ (Model 1)' and '$r_s$ (Model 2)'.

It should be noted that in the case of particle acceleration with
a broad energy distribution, most relativistic electrons may lose
energy very efficiently even if the average energy dissipation 
efficiency is low.  It means that, independently of the 
total energetics, the emissivity of such electrons
will be maximized very close to the shock front and its spatial 
distribution will match the distribution of the energy dissipation.
%Hence the knowledge of the shock front structure  is crucial to determine 
%Doppler boosting, relativistic abberation  effects, and polarization
%properties.

\section{Astrophysical appearance}

As theoretical analyses and numerical simulations demonstrate,
formation of reconfinement shocks is  accompanied by 
formation of reflection shocks (\citealt{1983ApJ...266...73S};
KF97). Furthermore, 
depending on a distribution of pressure or density of external medium, 
reconfinement and reflected shocks 
can form more or less abundant sequences of reconfinement shocks. 
Their radiative appearance is
commonly modeled by assuming proportionality of the emissivity to the gas pressure
(\eg~\citealt{2002LNP...589..169G}; KF97).
This leads to the predictions that most of the nonthermal radiation is produced
around the reflection shocks. However, proportionality of the emissivity to 
the pressure is not what should be expected, if the efficiency of particle 
acceleration scales with the efficiency of energy dissipation. 
The latter is maximized at the shock fronts and, 
therefore, radiation emitted by most relativistic electrons  will match
geometrical structure of the shock fronts rather than the volume distribution
of the pressure in the post shock flows. Of course, 'the shock front radiation'
is likely to be accompanied by emission from the entire post-shock volume,
by both slowly cooling lower energy electrons and by 
electrons accelerated in turbulent plasma in the $2^{nd}$ order Fermi process. 
Specific geometrical and kinematical structures of reconfinement shocks 
are expected to be reflected in polarization properties, provided that
magnetic fields are dominated by the shock compressed random field 
(\citealt{1980MNRAS.193..439L}; \citealt{1990ApJ...350..536C}).
This may explain perpendicular to the jet direction of 
polarization (EVPA) of radio knots 
in AGN kiloparsec scale jets \citep{1994AJ....108..766B}.
%Closer to the center, perpendicular or parallel EPV may be seen depending
%on which component dominates. One should note however that too strong
%toroidal component may suppress the shock formation (refs.).

\section{Conclusions}

\begin{itemize}
\item[-]
Semi-analytical models were developed to calculate the structure of
reconfinement shocks based on exact conservation laws and exact angular
relations. The approximate analytical formulae of KF97, that
describe a shape of the reconfinement shock and its dependence
on the power of a jet and the pressure of external medium, were confirmed 
with a very good accuracy, even for $\Gamma_j\Theta_j$ up to a few.
However, the absolute size of the structure is found to be larger
by a factor about two, when including the transverse pressure gradient
in the post-shock flow.

\item[-]
The efficiency of energy dissipation in the relativistic reconfinement shocks
scales approximately as $(\Gamma_j\Theta_j)^2$ for $\Gamma_j\Theta_j < 1$
and reaches about 6\% at $\Gamma_j\Theta_j = 1$. For both models, with or
without the transversal pressure gradients, the efficiency is very similar
and for a given value of $\Gamma_j\Theta_j$ practically does not depend on 
the bulk Lorentz factor.
\end{itemize}

\section*{Acknowledgments}

The present work was partially supported by the Polish Astroparticle Network
621/E-78/SN-0068/2007.

\bibliography{}

\appendix

\section{Solving the shock jump equations}
\label{appendix1}

We show here a method by which the parameters of matter behind the shock 
may be determined in an exact analytical manner from equations 
(\ref{eq1a} -- \ref{eq1d}). First, we express the angular parameters 
with non-angular ones. From equation (\ref{eq1a}) we find an 
expression for $\alpha_s$, which is also a differential equation for 
the shock surface:
\be
\label{app1a}
\tan\alpha_s=\der{r_s}{z}=-\frac{\beta_s\cos\theta_s-\beta_j\cos\theta_j}{\beta_s\sin\theta_s-\beta_j\sin\theta_j}
\ee
From equations (\ref{eq1b}) and (\ref{app1a}), we find the deflection angle of velocity field:
\be
\label{app1b}
\cos(\theta_s-\theta_j)=\frac{\Gamma_s\rho_s\beta_s^2+\Gamma_j\rho_j\beta_j^2}{(\Gamma_s\rho_s+\Gamma_j\rho_j)\beta_s\beta_j}
\ee
Now we find two more equations for two unknown parameters: $\Gamma_s$, $\rho_s$:
\be
\label{app1c}
\frac{w_s}{w_j}=\frac{\Gamma_j\rho_s}{\Gamma_s\rho_j}
\ee
\be
\label{app1d}
(\Gamma_s^2-\Gamma_j^2)\rho_sw_j=(p_j-p_s)\Gamma_s(\Gamma_s\rho_s+\Gamma_j\rho_j)
\ee
Equation (\ref{app1c}) is the result of dividing equation (\ref{eq1d}) by equation
(\ref{eq1b}). Equation (\ref{app1d}) is derived from equation (\ref{eq1c})
by eliminating trigonometric functions using equations (\ref{eq1a} -- \ref{eq1b})
and eliminating $w_s$ using equation (\ref{app1c}). Using the equation of state for
the shocked matter (and choosing the value of $\gamma_s$), we finally find from equations
(\ref{app1c}) and (\ref{app1d}) a quadratic equation for $\Gamma_s$:
\be
\label{app1e}
\renewcommand{\arraystretch}{1.4}
\setlength{\arraycolsep}{0pt}
\begin{array}{l}
\left[\gamma_sp_e(w_j-p_j+p_e)\right]\Gamma_s^2+\\
\qquad+\left[(\gamma_s-1)(p_j-p_e)\rho_jc^2\right]\Gamma_j\Gamma_s+\\
\qquad\qquad+\left[-w_j((\gamma_s-1)p_j+p_e)\right]\Gamma_j^2\;=\;0.
\end{array}
\ee
Analyzing the parameters of this equation, we know that there is always only
one positive solution. During our calculations, we set an alert for
unphysical $\Gamma_s<1$, but it never triggered.
Finding $\Gamma_s$, we calculate $\rho_s$ from equation
(\ref{app1d}) and then we find the angular parameters: $\theta_s$ from equation
(\ref{app1b}) and $\alpha_s$ from equation (\ref{app1a}).

\section{Solving the equations for conservation laws across the shocked zone}
\label{appendix2}

The normal vectors $\vec{n}_s$ and $\vec{n}_c$ are given explicitly by:
\bea
\vec{n}_s &=& -\cos\alpha_s\vec{e}_r+\sin\alpha_s\vec{e}_z,\\
\vec{n}_c &=& \cos\alpha_c\vec{e}_r-\sin\alpha_c\vec{e}_z.
\eea
Equations (\ref{eq3a} -- \ref{eq3b}) may be expanded into:
\be
\label{app2a}
\der{}{z}\left[\int_{r_s}^{r_c}u\rho\cos\theta r\;dr\right]=\left(u_s\rho_s\sin\delta_s\right)\frac{r_s}{\cos\alpha_s},
\ee
\be
\label{app2b}
\der{}{z}\left[\int_{r_s}^{r_c}\Gamma uw\cos\theta r\;dr\right]=\left(\Gamma_su_sw_s\sin\delta_s\right)\frac{r_s}{\cos\alpha_s},
\ee
\be
\label{app2c}
\renewcommand{\arraystretch}{1.4}
\setlength{\arraycolsep}{0pt}
\begin{array}{l}
\displaystyle
\der{}{z}\left[\int_{r_s}^{r_c}u^2w\sin\theta\cos\theta r\;dr\right]=\\
\displaystyle\qquad
=\left(u_s^2w_s\sin\delta_s\sin\theta_s+p_s\cos\alpha_s\right)\frac{r_s}{\cos\alpha_s}-p_er_c,
\end{array}
\ee
\be
\label{app2d}
\renewcommand{\arraystretch}{1.4}
\setlength{\arraycolsep}{0pt}
\begin{array}{l}
\displaystyle
\der{}{z}\left[\int_{r_s}^{r_c}\left(u^2w\cos^2\theta+p\right)r\;dr\right]=\\
\displaystyle\qquad
=\left(u_s^2w_s\sin\delta_s\cos\theta_s-p_s\sin\alpha_s\right)\frac{r_s}{\cos\alpha_s}+p_er_c\tan\alpha_c.
\end{array}
\ee

To perform the integrals we have to describe the flow parameters between $r_s$ and $r_c$ as functions of $r$. We notice that the expressions to be integrated are linear functions of $\rho$ and $p$ (since the enthalpy $w$ is also their linear function) and non-linear functions of $\Gamma$ and $\theta$. We decompose the integrated functions into $f(r)=g(\Gamma(r),\theta(r))\cdot h(r)\cdot r$, where $h(r)$ is one of $\rho(r)$, $p(r)$ or $w(r)$. We assume that $h(r)$ is linear:
\be
h(r)=h_s+\frac{h_c-h_s}{r_c-r_s}(r-r_s).
\ee
The integrals are approximated with
\be
\int_{r_s}^{r_c}f\;dr\simeq\frac{g(\Gamma_s,\theta_s)+g(\Gamma_c,\theta_c)}{2}\int_{r_s}^{r_c}h(r)r\;dr.
\ee
Substituting these formulae to equations (\ref{app2a} -- \ref{app2d})
we obtain a system of 4 differential equations for 8 variables:
$\Gamma_s$, $\rho_s$, $\theta_s$, $\alpha_s$, $p_s$, $\Gamma_c$, $\rho_c$ and $\alpha_c$.
The system is then closed by including differential forms of equations
(\ref{eq1a} -- \ref{eq1d}).

\end{document}